\pdfoutput=1
\documentclass[useAMS,usenatbib]{mnras}

\usepackage{graphicx} 
\usepackage{times}
\usepackage{epsfig}
\usepackage{epstopdf}
\usepackage{amsfonts}
\usepackage{amsmath}
\usepackage{amsbsy}
\usepackage{bm}
\usepackage{url}
\usepackage{microtype}
\usepackage{rotating}
\usepackage{sidecap}
\usepackage{longtable}
\usepackage{float}
\usepackage{lscape}
\usepackage{rotating}
\usepackage{amssymb,xcolor}
\include{def_bibtex}
\usepackage{booktabs,caption}
\usepackage[flushleft]{threeparttable}
\usepackage{soul}
\usepackage{colortbl}

\definecolor{amethyst}{rgb}{0.6, 0.4, 0.8}
\definecolor{blue-violet}{rgb}{0.54, 0.17, 0.89}

\def\ergcms{{\rm erg\,cm^{-2}\,s^{-1}}}

\usepackage[T1]{fontenc}
\usepackage{ae,aecompl}

\usepackage{newtxtext,newtxmath}

\title[X-ray dips and a CCF of MAXI J1820+070]{X-ray dips and a complex UV/X-ray cross-correlation function in the black hole candidate MAXI J1820+070}

\author[J.~J.~E. Kajava et al.]{J.~J.~E. Kajava,$^{1}$\thanks{E-mail: jari.kajava@utu.fi}
S.~E. Motta,$^{2}$
A. Sanna,$^{3}$
A. Veledina,$^{4, 5, 6}$
M. Del Santo$^{7}$\newauthor and 
A. Segreto$^{7}$ \\
$^{1}$Finnish Centre for Astronomy with ESO (FINCA), FI-20014 University of Turku, Finland\\
$^{2}$University of Oxford, Department of Physics, Astrophysics, Denys Wilkinson Building, Keble Road, Oxford OX1 3RH, UK\\
$^{3}$Dipartimento di Fisica, Universit\'a degli Studi di Cagliari, SP Monserrato-Sestu km 0.7, I-09042 Monserrato, Italy \\
$^{4}$Department of Physics and Astronomy, FI-20014 University of Turku, Finland\\
$^{5}$Nordita, KTH Royal Institute of Technology and Stockholm University, Roslagstullsbacken 23, SE-10691 Stockholm, Sweden\\
$^{6}$Space Research Institute of the Russian Academy of Sciences, Profsoyuznaya Str. 84/32, 117997 Moscow, Russia\\
$^{7}$ Istituto Nazionale di Astrofisica, IASF Palermo, Via U. La Malfa 153, 90146, Palermo, Italy\\
}
 
\date{}

\pubyear{2019}

\hypersetup{draft}

\begin{document}
\label{firstpage}
\pagerange{\pageref{firstpage}--\pageref{lastpage}}
\maketitle

\begin{abstract}
 MAXI~J1820+070, a black hole candidate first detected in early March 2018, was observed by \emph{XMM-Newton} during the outburst rise.
In this letter we report on the spectral and timing analysis of the \emph{XMM-Newton} X-ray and UV data, as well as contemporaneous X-ray data from the \emph{Swift} satellite.
The X-ray spectrum is well described by a hard thermal Comptonization continuum.
The \emph{XMM-Newton} X-ray light curve shows a pronounced dipping interval, and spectral analysis indicates that it is caused by a moderately ionized partial covering absorber.
The \emph{XMM-Newton}/OM $U$-filter data does not reveal any signs of the 17~hr orbital modulation that was seen later on during the outburst decay. 
The UV/X-ray cross correlation function shows a complex shape, with a peak at positive lags of about 4 seconds and a pre-cognition dip at negative lags, which is absent during the X-ray dipping episode. 
Such shape could arise if the UV emission comes partially from synchrotron self-Compton emission near the black hole, as well as from reprocessing of the X-rays in the colder accretion disc further out.

\end{abstract}




\section{Introduction}\label{sec:intro}

On March 11, 2018, MAXI/GSC triggered on a bright X-ray transient source MAXI~J1820+070 \citep{Kawamuro2018}.
The position of the transient was consistent with that of ASASSN-18ey, an optical transient discovered 5 days earlier \citep{Denisenko2018}.
The optical counterpart had also been detected in quiescence by \emph{Gaia} at coordinates RA = 18:20:21.94, DEC = +07:11:07.19 ($G = 17.41$ mag), with an estimated distance of $3.8^{+2.9}_{-1.2}$~kpc based on the measured parallax \citep{GRJ19}.
Rapid follow-up observations at various wavelengths all suggested that MAXI~J1820+070 is a black hole transient, initially detected in the hard spectral state (e.g., \citealt{Uttley2018}, \citealt{Bright2018}).
 
By March 14 MAXI~J1820+070 had brightened to 13th magnitude in the optical \citep{Baglio2018}. 
In the soft X-ray band \textsc{NICER} had seen evidence of dipping (as reported later in \citealt{Homan2018}), similar to those observed in high-inclination black hole X-ray binaries, such as GRO~J1655$-$40 and 4U~1630$-$47 \citep{KWB98,TLK98,KinZC2000,TUB03,TCG05,DTM14}. 
Such dips can be caused either by absorption in the bulge (or hot spot) where the matter stream from the companion star impacts the accretion disc (e.g., \citealt{WS82,AL98}), or by an equatorial wind launched from the disc (e.g., \citealt{Begelman1983,Ponti2012}). 
Given the high optical brightness and the presence of the X-ray dipping events, we triggered an \emph{XMM-Newton} (\emph{XMM}) target-of-opportunity observation of MAXI~J1820+070 on March 14, as signatures of the orbital period could potentially be detected in both \emph{XMM}/OM and \emph{XMM}/EPIC-PN data.
Moreover, simultaneous X-ray and UV data allow to shed light on the origin of the UV emission, by cross correlating the light curves in the two bands  \citep{Kanbach2001, HHC03}.
The \emph{XMM} observation was carried out on March 17-18, 2018 when the luminosity was still rising and the source was in the hard state. 
In this letter we report on the spectral and timing analysis of these \emph{XMM} X-ray and UV data, which we complement with simultaneous data from the Neil Gehrels \emph{Swift} observatory.

\section{\emph{XMM-Newton} and \emph{Swift} observations}

The \emph{XMM} observation consisted of 20~ks of EPIC-PN timing mode data, followed by 80~ks of burst mode data (grey and black sections, respectively, in Fig.~\ref{fig:XMM_curves}a). 
We processed the \emph{XMM} data using the Science Analysis Software (SAS) v. 16.0.0 with the up-to-date calibration files. 
We filtered events within the energy range 0.3--10.0~keV, retaining single and double pixel events only (pattern$\leq$4).
The \emph{XMM}/OM was operated in the fast mode with 1 second time resolution, using the $U$ filter.
As the \emph{XMM} X-ray light curve reveals clear dips, we extracted X-ray light curves at various energy bands to study them in detail.
We also extracted X-ray spectra outside these dips from the burst mode data, as well as an averaged dip spectrum by considering only count rates below 2500~cts~s$^{-1}$.
For the spectral extraction we selected single and double pixel events only (pattern$\leq$4) and we set `FLAG = 0', retaining events optimally calibrated in energy. We then extracted source and background spectra selecting events in the ranges RAWX=[30:44] and RAWX=[2:8], respectively. We applied energy rebin to the spectra such that we oversampled by a factor of 3 the energy resolution ensuring a minimum of at least 25 counts per bin. We investigated the `non-dipping' source spectrum for pile-up effects by comparing the spectrum extracted in the selection RAWX=[30:44] with the spectra extracted excising one, three, five and seven central columns of the aforementioned RAWX range. Comparing the evolution of the spectral properties as a function of the number of excised central columns, we found that pile-up effects are clearly mitigated removing five of the brightest central column of the EPIC-PN CCD.

\emph{Swift} also performed one pointed observation with the XRT instrument in windowed timing mode during the \emph{XMM} burst mode exposure, which is shown in Fig.~\ref{fig:XMM_curves}a in magenta colour.
In order to minimise the effects of pile-up, we extracted only grade 0 events from the \emph{Swift}/XRT data, and we generated a 0.6--10~keV energy spectrum considering only events collected in an annular region centred at the source position with inner and outer radius 6 and 20 pixels, respectively. 
We used the BAT-IMAGER software \citep{SCF10} which performs screening, mosaicking and source detection, to produce an averaged \emph{Swift}/BAT spectrum.
This spectrum was gathered from the survey data taken within the time interval of MJD~58195.00125--58196.00296, for a total of 13~ks exposure time.
The spectrum was extracted in 30 channels with logarithmic binning in the energy range of 15--185~keV. We used the standard BAT response matrix for the spectral modeling.
The resulting X-ray spectra were fitted with \textsc{xspec} v. 12.10.0, and all the errors are quoted at 1$\sigma$ level.

\section{Results}

\begin{table}
\centering
\caption{Best fitting parameters for the four models considered. The fluxes are given in units of $10^{-8}\,\ergcms$, and the disc normalization $K_\textrm{dbb}$ in $10^{5} \times (R_\textrm{in}\,[\textrm{km}]/d\,[\textrm{10\,kpc}])^2 \cos i$ units (where $R_\textrm{in}$ is the inner disc radius and $i$ is the inclination). Absorption columns, $N_\textrm{H}$, are in units of $10^{22}\,\textrm{cm}^{-2}$, the ionization parameter $\log (\xi)$ in (erg cm s$^{-1}$) and energies and temperatures are in keV. }\label{tab:parameters}
\begin{tabular}{l c c c c}

\hline
Parameter		& \multicolumn{2}{c}{Non-dipping spectrum} 				  & \multicolumn{2}{c}{Average dip spectrum}   				 \\
\hline
\hline		 

$N_\textrm{H, ISM}$ & $0.026_{-0.005}^{+0.005}$ & $0.14_{-0.03}^{+0.03}$ & [0.026]  & [0.14]  \\
$PCF_\textrm{1}$    & - & - & $0.38_{-0.04}^{+0.04}$    & $0.38_{-0.04}^{+0.04}$ \\
$N_\textrm{H1}$     & - & - & $16_{-3}^{+3}$    & $1.2_{-0.3}^{+0.3}$ \\
$PCF_\textrm{2}$    & - & - & $0.50_{-0.03}^{+0.02}$    & $0.55_{-0.04}^{+0.03}$ \\
$N_\textrm{H2}$     & - & - & $1.7_{-0.4}^{+0.4}$   & $13.7_{-1.5}^{+2}$ \\
$\log (\xi)$               & - & - & $0.9_{-0.2}^{+0.2}$ & $2.04_{-0.05}^{+0.05}$ \\
$kT_\textrm{dbb}$    & - & $0.200_{-0.010}^{+0.012}$  & -    & [0.200] \\
$K_\textrm{dbb}$    & - & $1.9_{-0.7}^{+1.2}$  & -      & [1.9] \\
$\Gamma$            & $1.630_{-0.009}^{+0.009}$ & $1.619_{-0.008}^{+0.010}$  & [1.630]  & [1.619] \\
$kT_\textrm{e}$      & $49_{-5}^{+6}$ & $44_{-3}^{+4}$  & [49]   & [44] \\
$\mathfrak{R}$       & $0.36_{-0.09}^{+0.10}$ & $0.27_{-0.08}^{+0.09}$  & [0.36]    & [0.27] \\
$E_{\rm Fe}$        & $6.68_{-0.06}^{+0.06}$ & $6.67_{-0.06}^{+0.06}$  & [6.68]     & [6.67] \\
$E_{\rm \sigma}$    & $0.92_{-0.09}^{+0.10}$ & $0.84_{-0.08}^{+0.09}$  & [0.92]     & [0.84]  \\
$F_{\rm bol}$       & $10.1_{-0.4}^{+0.3}$ & $9.9_{-0.4}^{+0.3}$  & -   & - \\
$F_{0.6-10}$        & $2.229_{-0.011}^{+0.011}$ & $ 2.226_{-0.030}^{+0.006}$  & $ 1.159_{-0.006}^{+0.011}$  & $1.152_{-0.006}^{+0.014}$  \\
$C_{\rm XMM}$       & $ 0.986_{-0.005}^{+0.005}$ & $ 0.987_{-0.007}^{+0.007}$  & $0.76_{-0.02}^{+0.02}$     & $0.775_{-0.015}^{+0.015}$ \\
\hline
$\chi^2$            & 747.5 & 708.8  & 161.2    & 169.5 \\
d.o.f.              & 742 & 740  & 154  & 154 \\
\hline
\end{tabular}
\end{table}

\begin{figure*}
\begin{tabular}{l r}
\includegraphics[width=0.56\textwidth]{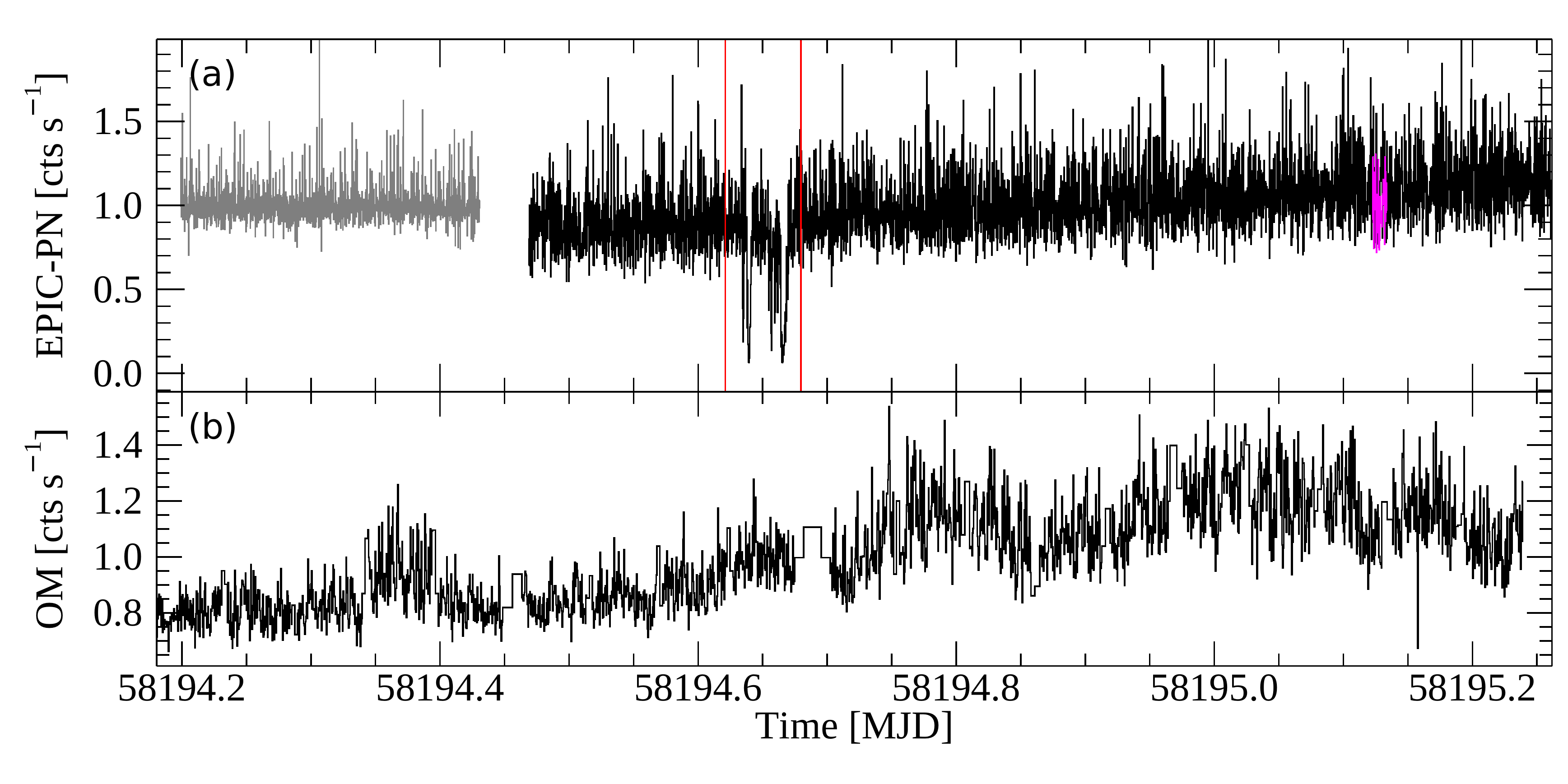} & \includegraphics[width=0.38\textwidth]{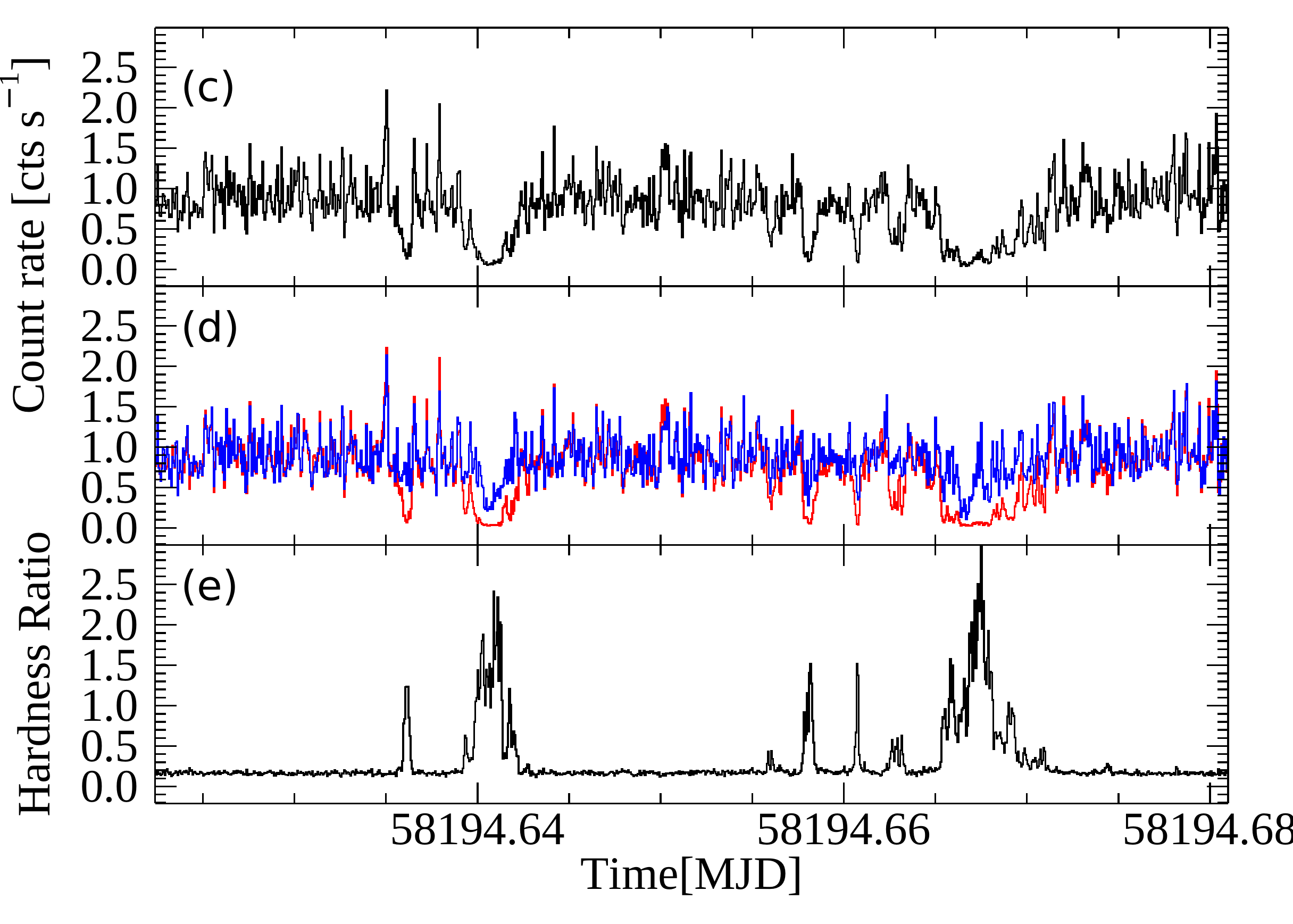}
\end{tabular}
\caption{Panel a: normalized \emph{XMM}/PN (timing mode in grey, burst mode in black) and \emph{Swift}/XRT light curves (magenta). Panel b: normalized \emph{XMM}/OM light curve. 
The averaged count rates for the burst mode data, the timing mode data, the \emph{Swift}/XRT data and the OM data are 7000, 1668, 522, 412 cts~s$^{-1}$, respectively.
Panel c: light curve of the dipping interval extracted in the 0.3--10~keV energy band and normalised to unity. 
Panel d: light curves in the 0.3--3.5~keV (red) and 3.5--10~keV (blue) energy bands, respectively. 
Panel e: the hardness ratio (3.5--10~keV)/(0.3--3.5~keV).
}
\label{fig:XMM_curves}
\end{figure*}

The X-ray light curve in Fig.~\ref{fig:XMM_curves}a shows a single dipping episode, which begins at MJD~58194.6345 and lasts until MJD~58194.6700, with a duration of $\approx$56~min. The simultaneous \emph{XMM}/OM UV data do not show any signs of dipping.
Since the dipping does not repeat, we can determine a minimum recurrence time of $\approx$15~hr and 6~min, measured  from the start of the first dip until the end of the \emph{XMM} observation.
A zoom-in of the dipping interval is shown in panels c--e of Fig.~\ref{fig:XMM_curves}.
A comparison between the light curves in the full 0.3--10~keV energy band (panel c) and the soft 0.3--3.5~keV (red) and the hard 3.5--10~keV (blue) energy bands (panel d), clearly indicate that the dipping is more pronouced in the soft band. 
From the hardness ratio between these two bands (panel e), we can see more clearly how the 2 major dips -- lasting $\approx$6 and $\approx$11~min -- are harder than the 5 shorter and shallower dips that have durations between 1 and 2~min.

\begin{figure}
\centering
\includegraphics[]{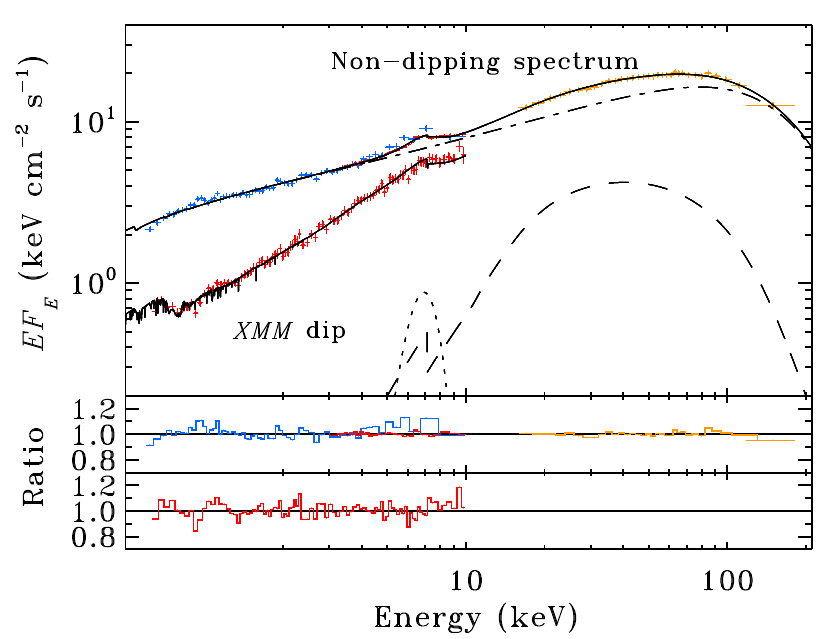}
\caption{The averaged X-ray spectrum of MAXI~J1820+070 during the \emph{XMM} dip (red), and outside the dip where the \emph{XMM}/PN, \emph{Swift}/XRT and BAT data are shown in red (not seen well on this scale), blue and orange points, respectively.
The dash-dotted, dotted and dashed lines show the \textsc{nthcomp}, \textsc{gauss} and \textsc{reflect} model components, respectively.
The bottom panels show the data/model ratios with respect to the non-dipping model and the dip model without the disc component.
}
\label{fig:spectra}
\end{figure}

An averaged X-ray spectrum of the dips, and the averaged ``non-dipping'' spectrum are shown in Fig.~\ref{fig:spectra}.
We used instrument cross-calibration constants in the fits, by fixing the \emph{Swift}/XRT constant to unity and allowed the \emph{XMM}/PN constant $C_{\rm XMM}$ and the \emph{Swift}/BAT constant to vary.
In the fitting we noted that below 3~keV the PN and XRT spectra are discrepant (likely due to calibration issues in the PN burst mode data) while above it they match nicely, and thus we fitted the PN spectrum in the 3--10~keV band and the XRT spectrum in the 0.6--10~keV band.
The ``non-dipping'' spectrum (see first column in Table~\ref{tab:parameters}) is well-described by an absorbed thermal Comptonization model \textsc{nthcomp} with a reflection component (\textsc{tbnew} $\times$ [\textsc{reflect} $\times$ \textsc{nthcomp} + \textsc{gauss}] model in \textsc{xspec}; see \citealt{Wilms2000,Magdziarz1995,Zycki1999}).
The best fitting electron temperature is about 50~keV, the photon index is $\Gamma \approx 1.63$, and the reflection amplitude $\mathfrak{R} \approx 0.36$, which are typical for a black hole in the hard state (e.g., \citealt{Gilfanov2010}).
The model fits the data well, giving $\chi^2$/d.o.f. $\approx 1$.

The addition of the \textsc{diskbb} model \citep{Mitsuda1984} (2nd column in Table~\ref{tab:parameters}) with $kT_\textrm{dbb} \approx 0.2$~keV improves the fit significantly ($\Delta \chi^2 = 38.7$ for two extra d.o.f.), although we note that the simpler model without the disc cannot be statistically rejected either. 
There is no appreciable change in the model parameters, except that the modelled interstellar absorption column increases five-fold from $N_\textrm{H, ISM} = 2.6 \times 10^{20}\,\textrm{cm}^{-2}$ to $N_\textrm{H, ISM} = 1.4 \times 10^{21}\,\textrm{cm}^{-2}$.
The best fitting disc normalization parameter $K_\textrm{dbb} \approx 1.9 \times 10^{5}\, (R_\textrm{in}\,[\textrm{km}]/d\,[\textrm{10\,kpc}])^2 \cos i$ can be used to estimate the apparent inner disc radius, yielding $R_\textrm{in} \approx 330$~km if we assume the distance $d \approx 3.8$~kpc and inclination $i \sim 60$ degrees, the latter value based on the presence of dipping and lack of X-ray eclipses.

In both of these models the broad ($\sigma_{\rm Fe} \approx 0.9$~keV) iron emission line is centered at $E_{\rm Fe} \approx 6.67$~keV, with an equivalent width of $EW_{\rm Fe} = 270 \pm 30$~eV.
We tested combinations of narrow iron emission and/or absorption lines at known energies of neutral iron (6.4~keV), or H-like (6.97~keV) and He-like ions (6.67~keV), but such models did not improve the fits significantly from the broad single Gaussian emission line that is reported in Table~\ref{tab:parameters}. Hence, we conclude that the data do not allow us to constrain any model more complex than those described above. 

In order to fit the dip spectrum, we assumed a priori that the persistent spectrum does not change appreciably from the average spectrum during the burst mode. 
We therefore fixed the continuum parameters to the ``non-dipping'' models, and studied how the dip spectrum can be described by multiplying the non-dipping spectrum with different absorption models (3rd and 4th columns in Table~\ref{tab:parameters}).
In these fits we allowed the \emph{XMM}/PN normalization constant $C_{\rm XMM}$ to vary, as the dip occurred in the beginning of the burst mode data when the X-ray flux was below the mean (see Fig.~\ref{fig:XMM_curves}a), and also the continuum level could in principle have changed during the dipping.
We found that the addition of simple absorption models did not provide good fits. 
For example, after adding a neutral partial covering absorber (\textsc{tbnew\_pcf}) to the model we obtain $\chi^2/\textrm{d.o.f.} = 1.63$, or $\chi^2/\textrm{d.o.f.} = 1.21$ if we add an ionized one (\textsc{zxipcf}).
We thus used a combination of these two models (\textsc{tbnew\_pcf} $\times$ \textsc{zxipcf}), which fits the data well with rather low neutral- and ionized absorber partial covering fractions of $PCF \approx 0.38 - 0.53$ and modest columns $N_\textrm{H} \approx [1.9 - 17] \times 10^{22}\,\textrm{cm}^{-2}$.
The best fitting ionization parameter was in the $\log(\xi) \approx 1.0-2.0$ range, depending on the chosen continuum model.
No significant absorption features were detected in the iron line region during the dipping.

\begin{figure}
\centering
\includegraphics[width=0.46\textwidth]{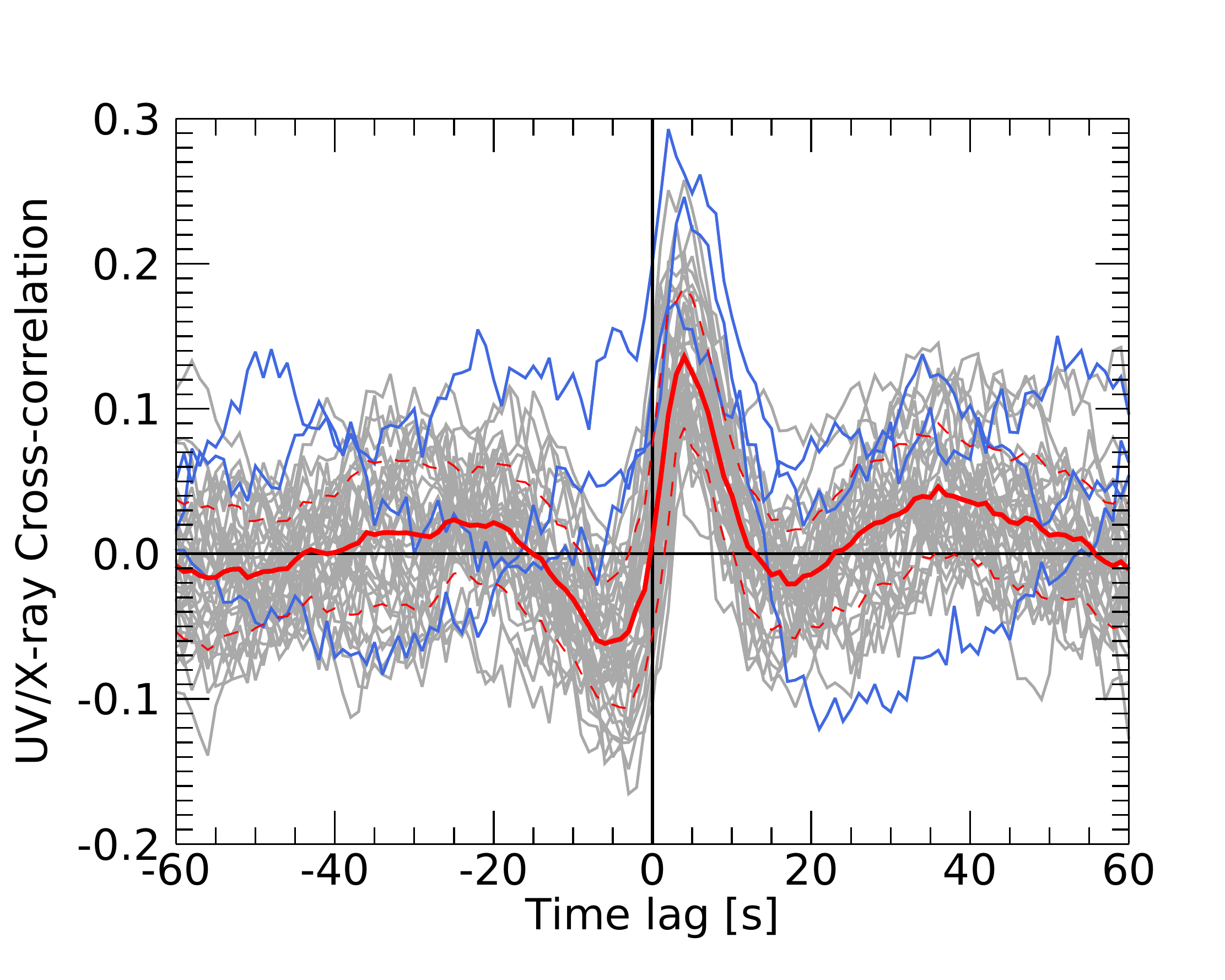}
\caption{UV/X-ray cross-correlation function from OM and PN burst mode data. Each grey line corresponds to a different OM window, while the red solid and dashed line show the mean and standard deviation, respectively. 
The three CCFs highlighted in blue correspond to the section of the light curves during the X-ray dips. 
}
\label{fig:CCFs}
\end{figure}

The almost uninterrupted day-long \emph{XMM}/OM $U$-filter data allowed us to search for periodicities that may have signatures of the binary orbital period.
However, a Lomb-Scargle periodogram did not reveal any significant periodic signal.
We instead used the \emph{XMM}/OM and the PN burst mode data to construct a series of cross-correlation functions (CCF). 
We extracted strictly simultaneous OM and PN light curves with the same time resolution of 1~sec, and computed CCFs for various window lengths (between a few hundreds and 2000~sec), after detrending each window with a linear function.
An optimal lenght of the window was found to be 1200~sec, which allowed to sample sufficiently short time-scales while maintaining a good S/N. 
The UV/X-ray CCFs for 36 of such windows are shown in Fig.~\ref{fig:CCFs} with grey lines, the 3 CCFs during dipping are highlighted in blue, while the average and standard deviation of the 36 CCFs are displayed with red solid and dashed lines, respectively. 
The 36 non-dipping CCFs are relatively stable during the entire observation.
For each window, the CCF has similar complex shape, with a clear dip at negative lags and a peak at positive time lags of $4.1\pm1.6$~sec.
The 3 CCFs taken during the dipping show no signs of the precognition anti-correlation. 

\section{Discussion}

X-ray dipping in MAXI~J1820+070 was also observed in the \textsc{NICER} data between March 12 and March 16, 
but it disappeared in the observations taken from March 21 onwards (\citealt{Homan2018}).
Moreover, the dipping became less and less pronounced from March 12 to March 16 (Homan et al. in prep.), and it is therefore possible that \emph{XMM} detected the last dipping episode from the source.
As the minimum recurrence time of 15~hr and 6~min is shorter than the likely orbital period of 17~hr \citep{Patterson2018ATel},
we cannot verify whether the subsequent dips were missed because they only appear at a given binary orbital phase.
The dipping behaviour of MAXI~J1820+070 resembles that of MAXI~J1659$-$152 \citep{Kuulkers2013}, which also showed dipping events in the outburst rise that ceased near the outburst peak and did not reappear in the ourburst decline.
Similarly to what we observe in MAXI~J1820+070, in MAXI~J1659$-$152 the longer, deeper dips were also harder than the shorter, shallow ones.
Interestingly, \citet{Munoz-Darias2018ATel} reported that the optical emission line profiles changed exactly during the same time period; prior to March 16 the Balmer and He-I line profiles had clear signatures of 1500 km s$^{-1}$ disc winds, which then disappeared later on. 
This suggests that the disc wind may play a role in generating the X-ray dipping we observe.

The broad band X-ray spectrum of the persistent, non-dipping emission measured by \emph{XMM} and \emph{Swift} is typical for a BH in the hard state.
Our best fitting values for the electron temperature and the photon index are consistent with those from black holes in the hard state (see, e.g., \citealt{Done2007}), and with the values obtained using the quasi-simultaneous MAXI and BAT data by \citet{SNY18}.
The dip spectrum required two additional partially covering absorbers to obtain a good fit, which is rather rare, though not unique.
For example, in GRO~J1655$-$40 the dips spectra can be fitted with a similar dual partial covering absorber model \citep{TUB03}.
Our best fitting ionization parameter $\log(\xi) \approx 1.0-2.0$ was similar to GX 339$-$4 (for which \citealt{MRF04} estimate $\log(\xi) \approx 1.8$) but lower than in several neutron star systems, which typically have $\log(\xi) \sim 2.2 - 3.9$ \citep{DiazTrigo2006}.
In these sources the ionization parameter tends to decrease during the dips while the absorption column increases.
Unfortunately we cannot study this behaviour with the \emph{XMM} data of MAXI~J1820+070, 
because of the relatively low S/N even in the averaged dip spectrum.
The low S/N also prevents us from finding evidence for narrow iron absorption features in the 6--8 keV range (such lines have been seen in GRO~J1655$-$40 and 4U~1630$-$47, see \citealt{DiazTrigo2007,DTM14}), particularly during the dipping.
Narrow absorption features are not either detected in the \textsc{NICER} observations taken from March 21 onwards \citep{KSF19}.

One should keep in mind that there are likely significant calibration issues in the \emph{XMM} burst mode data both in terms of flux and artificial residuals.
In the non-dipping spectrum where we have simultaneous \emph{Swift}/XRT data, the discrepancies between these two instruments reach factors of about 1.5 at 1~keV, 
while in the 3--10~keV range the spectra of the two instruments are consistent with each other.
While the dip spectrum should not be affected so severy by X-ray loading and pile-up given the lower count rate 
compared to the non-dipping spectrum, healthy scepticism and caution is in order when interpreting these fitting results. 
On the other hand, in several \textsc{NICER} observations taken prior to our \emph{XMM} pointing, evidence of mildly ionized absorbers were also seen (\citealt{Homan2018}, Homan et al. in prep.), lending support to our finding that the absorber is indeed mildly ionized.
The iron emission line seen in the \emph{XMM} data also points to an ionized reflector, the line energy being consistent with He-like Fe~\textsc{xxv}, which has the rest energy of 6.67~keV.
Similar broad iron lines with highly ionized species have been seen for example in XTE~J1748$-$288 \citep{MFdM01} as well as GX~339$-$4, which has broad lines primarily from H-like Fe~\textsc{xxvi} in the high-luminosity hard state \citep{PFP15}.
While a detailed modeling of the iron line is beyond the scope of this letter, we note that because the iron line seems ionized, the use of the neutral reflector model \textsc{reflect} is not entirely correct, since it generates a sharp edge at slightly incorrect energy.
Moreover, the broad line is inconsistent with the sharp edge in the \textsc{reflect} model, although blurring it with the \textsc{kdblur} model does not significantly change the reflection parameters.

The $kT_\textrm{dbb} \approx 0.2$~keV \textsc{diskbb} component would place the multicolour disc up to 780~km from the black hole (obtained assuming an inclination of 60$^{\circ}$), as the direct estimate should be multiplied by the square of the color-correction factor $\kappa \approx 1.7$ \citep{Kubota1998,Gierlinski1999,DBH05}.
Similar values were found by \citet{SNY18} in the hard state, but they are a factor of a few higher than seen later in the soft state \citep{Shidatsu2019}. 
The disc thus seemed to be truncated at fairly large radii ($\gtrsim20\,R_{\rm g}$), consistent with the low reflection amplitude of $\mathfrak{R} \approx 0.30$ we detect.
However, there are large uncertanties in the black hole mass, inclination and distance, and, in addition, the \textsc{diskbb} model normalization is strongly correlated with the column density, which makes this estimate quite speculative.
Moreover, the short reverberation lags measured in \citet{KSF19} suggest that the reflecting material is located close to the black hole, i.e., at $\sim14R_{\rm g}$ from the source of the incident X-ray continuum.
This points to either a very compact corona above the disc and the inner disc edge residing at the innermost stable circular orbit \citep{KSF19}, or to the fact that the source of the incident X-ray continuum is located close to the truncation radius, which in this case must have significantly shrank between the March 17 \emph{XMM} observation and the March 21 \textsc{NICER} one.

The complex UV/X-ray CCF can provide an independent test to see if the disc was truncated or not.
The observed complex shape of the UV/X-ray CCF is similar to the ones observed in XTE~J1118+480 \citep{HHC03}.
The precognition dip amplitude in those observations increases towards longer wavelengths, becoming most pronounced in the optical \citep{Kanbach2001}. 
This is also seen in MAXI~J1820+070 for which the dip in the optical/X-ray CCF as measured by \emph{Swift} XRT and UVOT is also broader and deeper than the \emph{XMM} UV/X-ray one \citep{Paice2018ATel}.
The presence of a precognition dip in the CCF can be interpreted as a signature of the synchrotron self-Compton mechanism operating in the hot accretion flow, similar to two other black hole binaries, Swift~J1753.5--0127 and BW~Cir \citep{Veledina2017,Pahari2017}.
Its essence is the increased synchrotron self-absorption in response to the increased mass accretion rate, which makes the X-rays anti-correlate with emission at longer wavelengths \citep{Veledina2011}.
The synchrotron emission from the accretion flow has a turnover at UV wavelengths if the flow size is $\sim20R_{\rm g}$ \citep{Veledina2013,Kajava2016}.
The covering of this region by the absorber during the X-ray dipping episode could have been responsible for the absence of the precognition dips in the CCFs.

The appearance of the positive peak can then be interpreted as the additional contribution of the delayed and smeared irradiated disc component, which has been seen also during the outburst peak in Swift~J1753.5--0127 \citep{Hynes2009}.
On the other hand, the IR/X-ray CCF of the prototypical black hole binary GX~339--4 shows a single positive peak at small, subsecond time-delays \citep{Casella2010}.
Its optical/X-ray CCF has a similar peak, but complemented by the precognition dip \citep{Gandhi2010}.
The single-peak CCF has been successfully modelled using the jet internal shocks scenario \citep[e.g.,][]{Malzac2018}, and the short delay in this model corresponds to the propagation time between the central engine and the region of maximal IR radiation within the jet.
The observed CCF in MAXI~J1820+070 peaks at time lags which are at least an order of magnitude higher than those detected in GX~339--4, and is more consistent with those of Swift~J1753.5--0127, thus supporting the irradiated disc scenario. 
This interpretation is also in line with the observed low polarization of optical $B$, $V$ and $R$-filter emission of MAXI~J1820+070 \citep{Veledina2019}, which was attributed to scattering processes in the disc atmosphere.

In summary, the single X-ray dipping episode detected in the \emph{XMM} observation points towards a long orbital period of $P_{\rm orb} \gtrsim 15$~hr for MAXI J1820+070. The dip spectrum can be modeled with a dual partially covering absorber with a modest ionization parameter. The non-dipping UV/X-ray CCF  -- with a pre-cognition dip and a positive peak at about 4 seconds -- is qualitatively similar to a few other black holes, which can be attributed to synchrotron-self-Compton emission in the hot flow near the black hole and X-ray reprocessing in the outer accretion disc. The fact that the pre-cognition dip disappears during the X-ray dipping suggests that the region responsible for the UV/X-ray anti-correlation is covered by the absorber. While providing only poor constraints if taken individually, considered together the morphology of the CCF, the low Compton reflection amplitude, and the disc component model parameters all suggest that the disc was at least mildly truncated during the outburst rise.

\section*{Acknowledgements}
JJEK acknowledges support from the Academy of Finland grant 295114.
SEM acknowledges support from the Violette and Samuel Glasstone Research Fellowship programme, the UK Science and Technology Facilities Council (STFC) and the FINCA visitor program.
ASa acknowledges support from the HERMES Project, financed by the Italian Space Agency (ASI) Agreement no. 2016/13 UO.
AV acknowledges support from the Academy of Finland grant 309308, and the Ministry of Science and Higher Education of the Russian Federation grant 14.W03.31.0021. 
SM and AV thank the International Space Science Institute (ISSI) in Bern, Switzerland for support.
MDS and ASe acknowledge financial contribution from the agreement ASI-INAF n.2017-14-H.0.
The authors thank the referee for suggestions that helped to improve the manuscript.




\bibliographystyle{mnras}
\bibliography{biblio} 

\label{lastpage}
\end{document}